\documentclass[aps,prb,twocolumn,superscriptaddress,showpacs]{revtex4-1}
\usepackage{amsmath}
\usepackage{graphicx}
\usepackage{epstopdf}
\usepackage{epsfig}
\usepackage{amsfonts}
\usepackage[breaklinks=true]{hyperref}
\usepackage{hypcap}
\usepackage{verbatim}
\usepackage{color}

\begin{document}

\title{Robust Two-Qubit Gates for Exchange-Coupled Qubits}

\author{F. Setiawan}
\author{Hoi-Yin Hui}
\affiliation{Condensed Matter Theory Center, Department of Physics, University of Maryland, College Park, Maryland 20742, USA}
\affiliation{Joint Quantum Institute, University of Maryland, College Park, Maryland 20742, USA}
\author{J. P. Kestner}
\affiliation{Department of Physics, University of Maryland Baltimore County, Baltimore, Maryland, 21250, USA}
\affiliation{Condensed Matter Theory Center, Department of Physics, University of Maryland, College Park, Maryland 20742, USA}
\author{Xin Wang}
\affiliation{Condensed Matter Theory Center, Department of Physics, University of Maryland, College Park, Maryland 20742, USA}
\author{S. Das Sarma}
\affiliation{Condensed Matter Theory Center, Department of Physics, University of Maryland, College Park, Maryland 20742, USA}
\affiliation{Joint Quantum Institute, University of Maryland, College Park, Maryland 20742, USA}
\date{\today}

\begin{abstract}
We present composite pulse sequences that perform fault-tolerant two-qubit gate operations on exchange-only quantum-dot spin qubits in various experimentally relevant geometries. We show how to perform dynamically corrected two-qubit gates in exchange-only systems with the leading hyperfine error term canceled. These pulse sequences are constructed to conform to the realistic experimental constraint of strictly non-negative couplings. We establish that our proposed pulse sequences lead to several orders of magnitude improvement in the gate fidelity compared with their uncorrected counterparts. Together with single-qubit dynamically corrected gates, our results enable noise-resistant universal quantum operations with exchange-only qubits.
\end{abstract}


\maketitle
\def\beq{\begin{equation}}
\def\eeq{\end{equation}}
\newcommand{\abs}[1]{\left|#1\right|}
\newcommand{\ket}[1]{\left|#1\right\rangle}
\newcommand{\bra}[1]{\left<#1\right|}
\newcommand{\dis}{\displaystyle}
\newcommand{\mathsc}[1]{\text{\textsc{#1}}}

\section{Introduction}
The Heisenberg exchange interaction \cite{Scarola} has been at the heart of spin qubit control since the introduction of spin-based quantum computation in semiconductor quantum dots.  The interdot exchange interaction provides the two-qubit gate when each spin is regarded as a qubit or, alternatively, if a qubit is encoded in the singlet and triplet two-spin states \cite{Levy, Petta, Bluhm, Shulman, Maune}, it constitutes a subset of single-qubit operations. If one qubit is encoded in three physical spins, the exchange interaction alone suffices for universal quantum computation \cite{Divincenzo}. This ``exchange-only'' qubit has the advantage of fast and all-electrical control and has therefore received extensive attention both theoretically \cite{West, Fong, Mehl,Pal.13} and experimentally \cite{Gaudreau1, Gaudreau2, Laird, Takakura, Gaudreau3, Hsieh, Medford, Braakman}. Very recently, a variation, the resonant-exchange qubit, has been experimentally demonstrated \cite{Medford.13, Jake, doherty}.

The universality of the exchange interaction comes with a cost in that entangling operations on such qubits involve the interqubit exchange coupling, which inherently causes leakage out of the logical subspace, necessitating additional gate operations. The leakage may be reduced by using a single adiabatic pulse \cite{doherty} or completely eliminated by performing an appropriate sequence of gates operations \cite{Divincenzo,Fong,West}. For adiabatic coupling, the geometry of the exchange-only qubit network plays an important role \cite{doherty}, while for the pulse-sequence approach, only a few geometries have been considered and most attention has been restricted to the linear spin chain.  Other geometries of current experimental interest, such as the ``butterfly" form \cite{doherty}, have no known leakage-correcting sequences. Furthermore, in both single-pulse and pulse-sequence approaches, decoherence through the hyperfine interaction with the nuclear spin bath (i.e., Overhauser fluctuations) remains an outstanding challenge for the exchange-only qubit. While at the single-qubit level that problem has recently been addressed using a form of dynamically corrected gates \cite{Hickman}, construction of a noise-resistant two-qubit exchange-only gate is a completely unexplored territory.  Thus, despite the great promise of the exchange-only spin qubit system, the basic problems of multiqubit gating must be addressed for rapid experimental progress to continue toward the construction of a fault-tolerant spin-based semiconductor quantum computer.

In this paper, we address these problems in two steps. First, we develop two-qubit pulse sequences for several different relevant geometries to eliminate the leakage intrinsic to interqubit coupling in exchange-only qubits even in the absence of noise. Second, based on these results, we develop a theoretical framework for implementing the two-qubit gate robust to leading-order quasistatic hyperfine noise. As Overhauser fluctuations associated with hyperfine noise are the dominant sources of error in GaAs \cite{Medford, Mehl}, our pulse sequences are relevant to ongoing experiments.

\section{Basis States and Operations}
An exchange-only qubit is encoded in the $S = 1/2$, $S^{z} = 1/2$ subspace of three-spin states as $|0\rangle = (\left|\uparrow \downarrow \uparrow \rangle\right. - \left|\downarrow \uparrow \uparrow \rangle\right.)/\sqrt{2}$ and $|1\rangle = ({\left|\uparrow \downarrow \uparrow\rangle\right.} + \left|\downarrow \uparrow \uparrow \rangle\right.)/\sqrt{6} - \sqrt{6}\left|\uparrow \uparrow \downarrow\rangle\right./3$ (see Ref.~\onlinecite{Divincenzo}). The four two-qubit logical states, together with the leakage states, live inside the nine-dimensional $S_{\mathrm{tot}} = 1$, $S^{z}_{\mathrm{tot}} = 1$ subspace of six-spin states. Neighboring spins ${\bf{S}}_i$ and ${\bf{S}}_j$ interact via the Hamiltonian $H^{\mathrm{ex}}_{ij}(t) = J_{ij}(t)E_{ij}$, where $E_{ij} \equiv {\bf{S}}_i\cdot{\bf{S}}_j$ and $J_{ij}(t)$ is the Heisenberg exchange interaction. $J_{ij}$ can be controlled electrostatically by changing gate voltages to adjust the interdot detunings. The exchange Hamiltonian generates a rotation $R_{ij}(\phi) \equiv \mathrm{exp} \left(-\imath E_{ij}\phi\right) = \mathrm{exp}[-\imath \int \mathrm{d}t H^{\mathrm{ex}}_{ij}(t)] $ about axis $ij$ in the nine-dimensional subspace (we set $\hbar = 1$ in this paper). A full {\sc swap} operation between spins $i$ and $j$ therefore corresponds to $R_{ij}(\pi)$, up to a global phase. We emphasize that, physically, the exchange coupling and pulse duration must always be non-negative, so $J_{ij}, \phi \geq 0$ [e.g., $R_{ij}(-\pi/2)$ must be carried out as $R_{ij}(3\pi/2)$].

Our purpose is to build a two-qubit gate using exchange pulses $R_{ij}(\phi)$ as the building blocks. Before we proceed, however, we must distinguish between two kinds of leakages. As mentioned above, in the absence of noise, a pulse using the interqubit exchange coupling inevitably introduces leakage. To understand this, consider the state $\left|\uparrow\uparrow\downarrow\right\rangle \otimes \left|\uparrow\uparrow\downarrow\right\rangle$. A {\sc swap} operation, between the rightmost electron of the first qubit and the leftmost electron of the second qubit, would leave the first qubit in $S^z=3/2$ and the second in $S^z=-1/2$ subspace, none of which is in the computational subspace.  Since a two-qubit gate necessarily utilizes the inter-qubit exchange pulse, we call this ``intrinsic leakage." We emphasize here that, because the intrinsic leakage is completely unrelated to noise, it can be fully compensated by combining various intraqubit and interqubit exchange pulses, as has been explicitly shown for the linear geometry \cite{Divincenzo,Fong}. On the other hand, even if intrinsic leakage is compensated, the hyperfine coupling to the nuclear spin bath causes additional leakage as well as dephasing, which we refer to collectively as hyperfine error. In what follows, we shall first present pulse sequences to compensate for the intrinsic leakage. Then we discuss how one may combat the hyperfine error by extending these sequences. For successful quantum computing operations, both errors must be suppressed below the error correction threshold.

\section{CNOT Sequences Free From Intrinsic Leakage}\label{sec:intrinsic}
In this work, we consider various geometries of exchange-coupled quantum-dot networks. For each geometry we obtain the corresponding sequence of unitary exchange interactions, the product of which, $W = \prod_k{R_{ij}^{(k)}\left(\phi_{ij}^{(k)}\right)}$, yields a {\sc cnot} gate (up to local unitaries) with the intrinsic leakage completely compensated. A genetic algorithm method has previously been used for this sort of task \cite{Fong}, but here we use a faster constrained exhaustive search method (see Appendix \ref{sec:algorithm} for details). The use of this method is motivated by the observation that the {\sc cnot} pulse sequences (excluding the local unitaries) presented in Refs. \onlinecite{Fong, zhan} consist of only pulses that are products of $\sqrt{\mathsc{swap}}$. Restricting ourselves to this kind of pulse, we carry out the search by calculating the two invariants $G_{1,2}(W)$ identified by Makhlin \cite{Makhlin}. Additional remarks about the optimization of the procedure can be found in Appendix B.

\begin{figure}[htb]
\centering
{\includegraphics[width = 1\columnwidth]{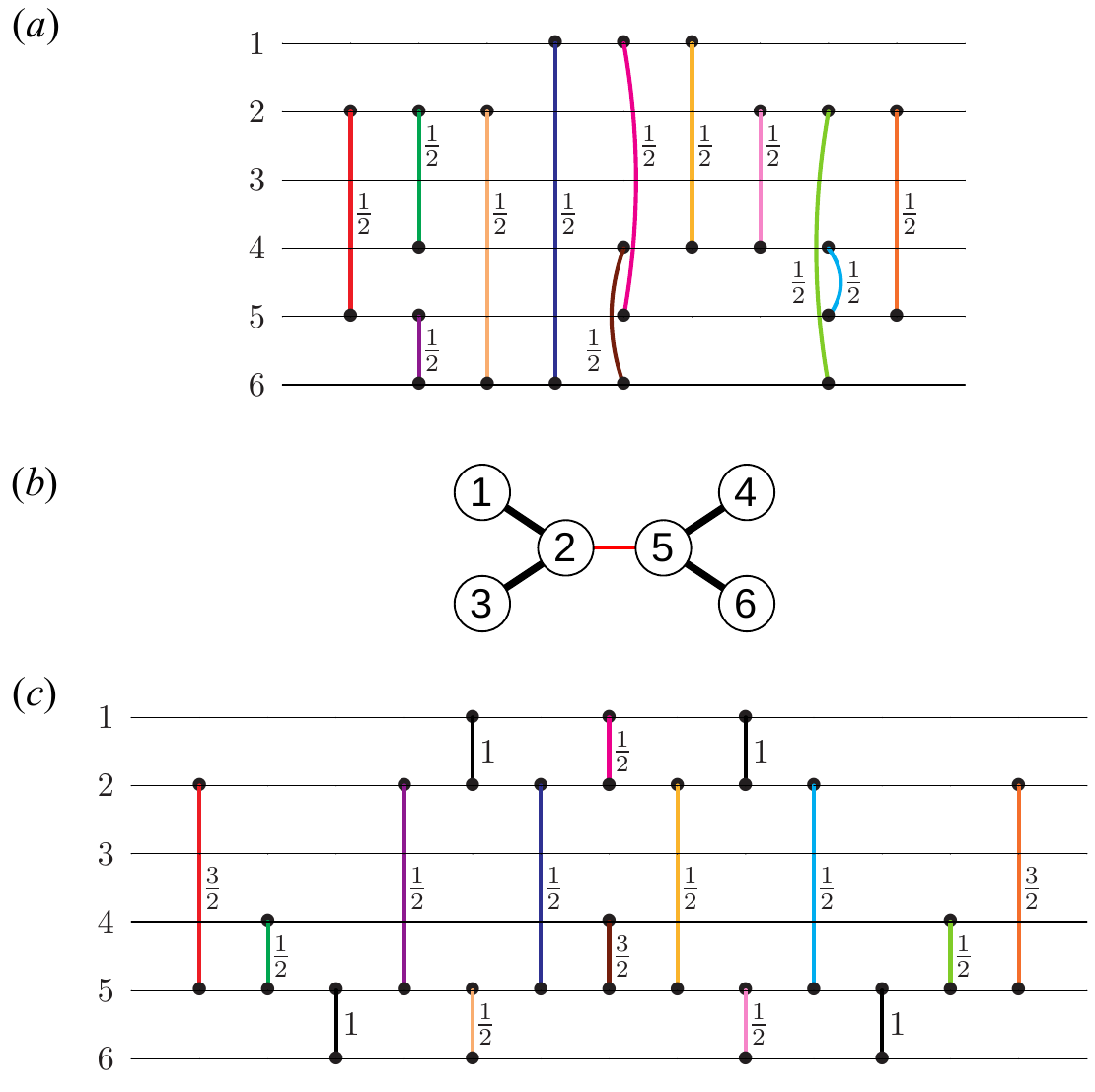}}
\caption{(Color online) (a) A {\sc cnot} sequence, consisting of 12 $\sqrt{\mathsc{swap}}$ pulses in 9 time steps, for a fully connected network. Note that at the fifth time step, the exchange pulses for links  15 and 46 are to be carried out simultaneously. The first qubit is encoded in spins 1, 2, and 3 while the second is encoded in spins 4, 5, and 6. (b) Schematic diagram of the butterfly geometry. (c) Schematic diagram of 16 exchange pulses in 13 time steps that realize a {\sc cnot} gate up to local unitaries for the butterfly geometry.}
\label{fig:butterfly}
\end{figure}

The shortest {\sc cnot} sequence that we find by this method consists of twelve $\sqrt{\mathsc{swap}}$ pulses in nine time steps, which is for a fully connected network. In this network, every pair of dots is linked, resulting in the sequence shown in Fig.~\ref{fig:butterfly}(a), where the time ordering is shown from left to right. Each pulse is labeled by its time duration $t$, given in units such that for $t=1$, $\mathrm{exp}(-\imath t J_{ij} {\bf{S}}_i\cdot{\bf{S}}_j)$  corresponds to a {\sc swap} operation between spins $i$ and $j$. The sequence shown is unique up to a relabeling of the dots. Related leakage-free sequences for other geometries can be obtained from the fully connected sequence by relabeling the dots as necessary and inserting additional full {\sc{swap}} operations to shuttle spins from unconnected dots to connected ones \cite{Ladd}.

For the linear geometry, we reproduce the result of Fong and Wandzura \cite{Fong}(see Appendix \ref{sec:linear}) and confirm that it is the optimal one under the current search constraints. Moreover, our method gives rise to shorter pulse sequences (see Appendix \ref{sec:linear}) for the geometries considered in Ref.~\onlinecite{zhan} and new pulse sequences for geometries whose two-qubit gate sequences have not been considered previously (see Appendix \ref{sec:linear}). As an example, we show our result for the butterfly geome

try in Fig.~\ref{fig:butterfly}(c). Our {\sc cnot} sequence (up to local unitaries) consists of 16 exchange pulses in 13 time steps. This sequence is equivalent to the sequence for a fully connected geometry shown in Fig.~\ref{fig:butterfly}(a). The mapping is shown by the color coding in Figs.~\ref{fig:butterfly}(a) and \ref{fig:butterfly}(c) in which equivalent $\sqrt{\mathsc{swap}}$ pulses are drawn by the same colors. For instance, the first $\sqrt{\mathsc{swap}}$ pulse is shown in red, the second is colored in green, and so on. The $\sqrt{\mathsc{swap}}$ between electrons in dots 2 and 4 in the second time step for the fully connected geometry sequence is implemented in the butterfly geometry by swapping electrons in dots 2 and 5 and then performing a $\sqrt{\mathsc{swap}}$ between electrons in dots 4 and 5. (We note that for the butterfly geometry, there exists an alternative single-pulse method of suppressing intrinsic leakage \cite{doherty}. See Appendix \ref{sec:doherty} for a comparison.)

\section{Hyperfine Noise Correction Scheme}\label{sec:hyperfine}
Having compensated completely the intrinsic leakage, we now consider the case where the exchange-only qubits are subject to Overhauser noise, which is the dominant noise in the experiment\cite{Medford, Mehl}. The hyperfine Hamiltonian is given by $H^\mathrm{hf}_{\mathrm{tot}} = \sum_{i=1}^{6}H_i^\mathrm{hf}$, where $H_i^\mathrm{hf}=E^{\mathrm{Z}}_i S_i^z$ and $E^{\mathrm{Z}}_i = g\mu_BB_i$. Here $g$ is the electron $g$-factor, $\mu_B$ is the Bohr magneton, $B_i$ is the random quasistatic Overhauser field at dot $i$, and $S^z_i$ denotes the $z$ component of the spin operator at that dot. An exchange pulse about the $ij$ axis evolves the state of a pair of spins $i$ and $j$ according to
\begin{align}
U_{ij}(J,\phi) = \mathrm{exp}\left[-\imath (JE_{ij}+H^{\mathrm{hf}}_{i}+H^{\mathrm{hf}}_{j})\frac{\phi}{J}\right].
\end{align}
For the convenience of later discussion, we also define a free evolution of a spin pair $ij$ for time $\tau$ when $J_{ij}=0$ as
\begin{align}
U_{ij}^{\mathrm{free}}(\tau) = \mathrm{exp}\left[-\imath \left(H^{\mathrm{hf}}_{i}+H^{\mathrm{hf}}_{j}\right)\tau\right].\label{eq:freeevol}
\end{align}

\begin{figure}[h!]
\includegraphics[width = 1\columnwidth]{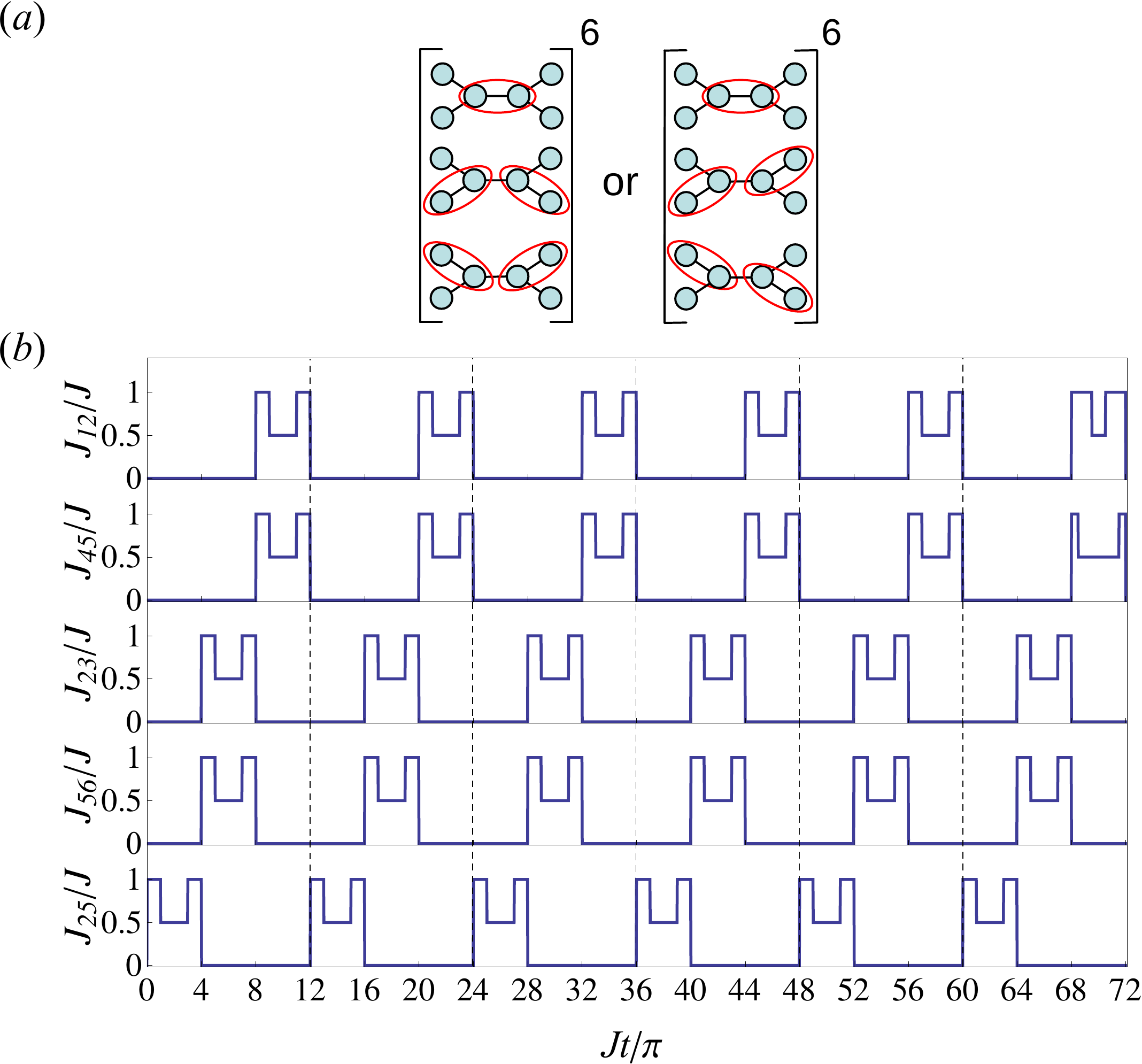}
\caption{(Color online) (a) Schematic depiction of the complete permutation cycle used for hyperfine
noise correction. Red ovals denote {\sc swap} operations. (b) Example of a pulse sequence implementing a corrected simultaneous rotation about axes 12 and 45 [Eq.~\eqref{eq:butterfly}] for the butterfly geometry. Here, we take $\phi_{12} = \pi/2$, and $\phi_{45} = -\pi/2$. The dashed lines separate the pulse sequence into blocks corresponding to the left diagram in (a), but with the last pulse modified according to Eq.~\eqref{eq:butterfly}.}\label{fig:EvolJ}
\end{figure}

We now describe the basic exchange pulse sequence to suppress the Overhauser noise.  The basic idea is to apply {\sc swap} operations ($\pi$ rotations) between pairs of neighboring dots such that each electron spends an equal amount of time on each dot \cite{West}. By doing so, an effective global system-bath interaction is generated so that the hyperfine error is turned into a global phase. To be concrete, in the following we present the results for the butterfly geometry shown in Fig. \ref{fig:butterfly}. For this geometry, the exchange-coupling terms can be grouped into three sets and within each set the elements commute. One such grouping is given by $E_A =\{E_{12},E_{45}\}$, $E_B = \{E_{23},E_{56}\}$, and $E_C = \{E_{25}\}$ [see Fig.~\ref{fig:EvolJ}(a)]. Simultaneously turning on the exchange interactions in each individual set results in exchange pulses $U_A(J,\phi) \equiv U_{12}(J,\phi)U_{45}(J,\phi)U^{\mathrm{free}}_{36}(\phi/J)$, $U_B(J,\phi) \equiv U_{23}(J,\phi)U_{56}(J,\phi)U^{\mathrm{free}}_{14}(\phi/J)$, \text{and} $U_C(J,\phi) \equiv U_{25}(J,\phi)U^{\mathrm{free}}_{14}(\phi/J)U^{\mathrm{free}}_{36}(\phi/J)$, where again $U^{\mathrm{free}}_{ij}(\phi/J)$ denotes the free evolution of spins on unconnected dots $i$ and $j$ while other couplings are turned on for a time $\phi/J$. A complete cycle of permutations can be performed by using the composite pulse $[U_A(J,\pi)U_B(J,\pi)U_C(J,\pi)]^6$ as depicted in the left diagram of Fig.~\ref{fig:EvolJ}(a) or its cyclic permutation counterparts $[U_B(J,\pi)U_C(J,\pi)U_A(J,\pi)]^6$ and $[U_C(J,\pi)U_A(J,\pi)U_B(J,\pi)]^6$. Note that the square brackets in Fig. \ref{fig:EvolJ}(a) should be understood as a product of several terms, each of which is represented by one graph, where an oval encircling spins $i$ and $j$ indicates a rotation $U_{ij}(J_{ij},\pi)$, while the remaining pairs, if any, should be understood as performing free evolution subject to hyperfine noise according to Eq.~\eqref{eq:freeevol}. The exponent on top of the square bracket indicates that these three steps are repeated six times to form the entire sequence. An alternative way of performing the permutation is shown in the right diagram of Fig. \ref{fig:EvolJ}(a).

Now, if the pulses $U_{ij}(J,\pi)$ are replaced by the composite pulses $U'_{ij}(J,\pi)$ \cite{Hickman}, where
\begin{align}
U'_{ij}(J,\phi) = U_{ij}(J, \phi)U_{ij}\left(\frac{J}{2},2\pi-\phi\right)U_{ij}(J,\phi),\label{eq:threepiece}
\end{align} and, since the time duration of the composite pulse is $4\pi/J$ instead of $\pi/J$, $U_{ij}^{\mathrm{free}}(\pi/J)$ is replaced by $U_{ij}^{\mathrm{free}}(4\pi/J)$,  then the complete cycle of permutations generates an identity with the first-order hyperfine error turned into a harmless global phase
\begin{align}
[U'_{A}&(J,\pi)U'_{B}(J,\pi)U'_{C}(J,\pi)]^6 \nonumber\\
&  = \exp\left[-\imath\left(\frac{\pi}{2}+12\pi\sum_{i=1}^6\frac{E^{\mathrm{Z}}_i}{J}\right)\right]I +O((E^{\mathrm{Z}}_i/J)^2),\label{eq:idbutterfly}
\end{align}
where the primes indicate the substitution mentioned. Note that for simplicity we have assumed boxcar pulses for $U'_{ij}(J,\phi)$, but finite rise and fall times can easily be incorporated by replacing the three-piece pulse $U'_{ij}(J,\phi)$ by the trapezoidal pulses given in Ref.~\onlinecite{Hickman}. Furthermore, we have also assumed the same value of $J$ for $U'_{A}(J,\phi)$, $U'_{B}(J,\phi)$, and $U'_{C}(J,\phi)$. However, in general, different $J$ values can be used for these three composite pulses.

A corrected nontrivial rotation, e.g., rotation about axis 12 [such as the second time step of the sequence in Fig.~\ref{fig:butterfly}(b)], can now be constructed by replacing the final segment of the identity sequence of Eq.~\eqref{eq:idbutterfly}, $U'_{12}(J,\pi)$, by $U'_{12}(J,\pi+\phi_{12})$ (for  $-\pi\leq\phi_{12}\leq\pi$). Similarly, a corrected rotation about any other axis $ij$ can be obtained by cyclically permuting the identity sequence such that its last segment is $U'_{ij}(J,\pi)$ and replacing that with $U'_{ij}(J,\pi+\phi_{ij})$.  Crucially, the error still cancels because $U'_{ij}(J,\phi)$ in Eq.~\eqref{eq:threepiece} was chosen such that the error is independent of $\phi$.

Note that although most of the steps in Fig.~\ref{fig:butterfly}(b) are single-qubit rotations, we cannot simply replace them by the shorter sequences developed in Ref.~\onlinecite{Hickman}. The sequence in Fig.~\ref{fig:butterfly}(b) is designed to compensate the intrinsic leakage, guiding the evolution through the leakage subspace and back into the logical subspace. At each step we must correct the error not only for states in logical subspace $(S^z = 1/2)$, but also for states in the leakage subspace $(S^z = -1/2$ or $3/2)$. The shorter sequences in Ref.~\onlinecite{Hickman} work only if the state always stays in the subspace with a particular value of $S^z$, so we must use the longer permutation sequence presented in this paper. We remark, however, that outside the two-qubit sequences [such as Fig.~\ref{fig:butterfly}(b)], the shorter sequences of Ref.~\onlinecite{Hickman} can be used for single-qubit gates since the leakage states play no role there.

At certain time steps in Fig.~\ref{fig:butterfly}(b), for example, the seventh step, two rotations that commute can be performed simultaneously. There the corrected simultaneous rotation about axes 12 and 45 can be performed by simply replacing the last composite pulse in the complete permutation sequence with $U'_{12}(J,\pi+\phi_{12})U'_{45}(J,\pi+\phi_{45})U^{\mathrm{free}}_{36}(4\pi/J)$ giving the total sequence
\begin{multline}
U'_{12}(\pi+\phi_{12})U'_{45}(\pi+\phi_{45})U^{\mathrm{free}}_{36}(4\pi/J) U_{B}'(J,\pi)U_{C}'(J,\pi)\\
\times [U'_A(J,\pi)U'_B(J,\pi)U_{C}'(J,\pi)]^5 \\
= R_{12}(\phi_{12}) R_{45}(\phi_{45})+O((E^{\mathrm{Z}}_i/J)^2)\label{eq:butterfly}
\end{multline}
up to a global phase. The time profile of the exchange couplings is then as shown in Fig.~\ref{fig:EvolJ}(b).

With these results it is then straightforward to replace every pulse shown in Fig.~\ref{fig:butterfly}(b) and make it robust against hyperfine errors. The infidelity of the {\sc cnot} sequence vs the strength of the random Overhauser field gradient is given in Fig.~\ref{fig:infidelity}. (The formula used to calculate the fidelity is given in Appendix \ref{sec:fidel}.) There for plotting purposes we have assumed that the Overhauser field difference between each dot and its neighbor(s) to the right is $\Delta E^{\mathrm{Z}}$, but that assumption is certainly not necessary for our pulse sequences to work. For comparison, we also show results for linear and loop geometries. The linear geometry is the most common in the literature \cite{Divincenzo, Fong, Pal.13} and the loop geometry is simply given for comparison as a natural way to add one link to the linear geometry. The sequences compensating intrinsic leakage for those two geometries are given in Ref.~\onlinecite{Fong} and can also be obtained from Fig.~\ref{fig:butterfly}(a) by adding {\sc swap} operations to account for the missing links. Furthermore, the noise correction scheme developed above for the hyperfine noise suppression in the butterfly geometry also applies for linear and loop geometries but with different sets of commuting exchange couplings, shown in Fig.~\ref{fig:permutation}. (Explicit formulas for these sequences are given in Appendix \ref{sec:linear}). Note that the CNOT sequences and noise-correction schemes given in this paper also work in the $S_{\mathrm{tot}} = 0$, $S^{z}_{\mathrm{tot}} = 0$ subspace of six-spin states \cite{Fong}.

\begin{figure}[h!]
\includegraphics[width=1\columnwidth]{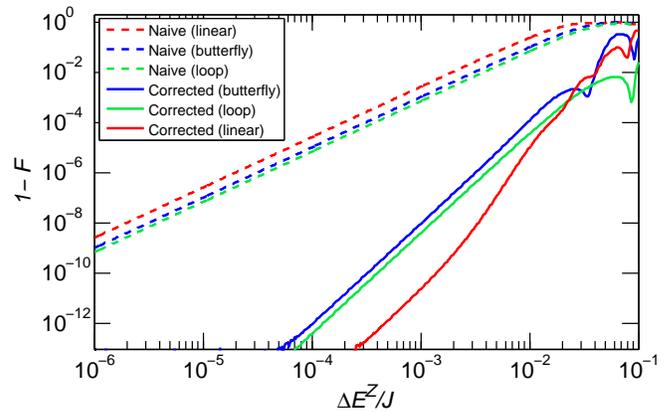}
\caption{(Color online) Gate infidelity of naive (dashed curves) and corrected (solid curves) {\sc cnot} pulse sequences vs magnetic field gradient \textit{$\Delta E^{\mathrm{Z}}/J$} for the butterfly (blue), linear (red) and loop (green) geometries.}
\label{fig:infidelity}
\end{figure}

\begin{figure}
\begin{center}
\includegraphics[scale = 0.6]{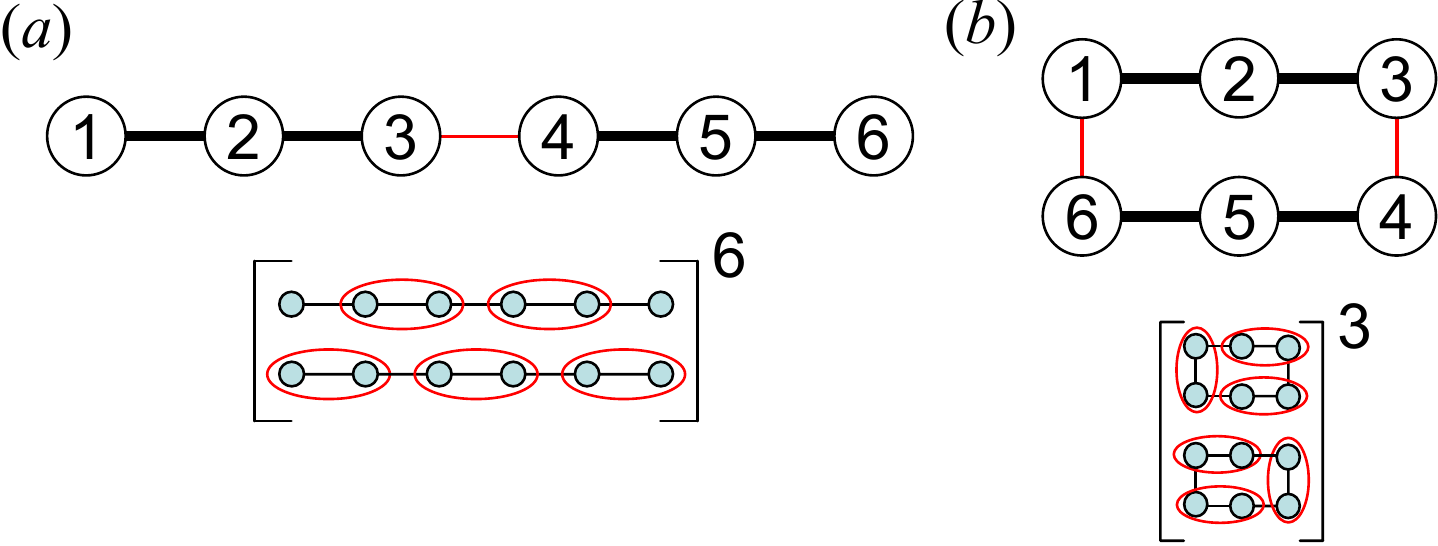}
\end{center}
\caption{(Color online) Permutation schemes to correct for the leading term of the hyperfine error for (a) linear and (b) loop geometries. Red ovals denote {\sc swap} operations.}\label{fig:permutation}
\end{figure}

As shown in Fig.~\ref{fig:infidelity}, for $\Delta E^{\mathrm{Z}}/J \ll 1$, the infidelities of the naive and corrected gates scale as $O((\Delta E^{\mathrm{Z}}/J)^2)$ and $O((\Delta E^{\mathrm{Z}}/J)^4)$, respectively. This reduced error rate is not without cost.  For every time step in Fig.~\ref{fig:butterfly}(b), the hyperfine-corrected pulse is  a $54 \pi$ rotation. The length of the entire corrected sequence is then $\sim700 \pi$, which is more than an order of magnitude longer than the uncorrected sequence.  It is possible that further optimization of our sequences might be obtained by switching on the noncommuting exchange couplings simultaneously, as in the resonant exchange qubit \cite{Medford}. However, the cost of using the corrected sequence can be justified by the tremendous improvement in the gate fidelity (as detailed below). Experiments in GaAs report slowly fluctuating hyperfine gradients with a standard deviation of $\Delta B \sim 3$ mT \cite{Medford}, corresponding to $\Delta E^{\mathrm{Z}} \sim 100$ neV. In that case, our sequence is useful for exchange couplings on the order of a $\mu$eV or more.  Reference \onlinecite{Medford} reports exchange values up to $30$ $\mu$eV ($\Delta E^{\mathrm{Z}}/J \sim 3 \times 10^{-3}$), at which point our corrected {\sc cnot} pulse sequence for the butterfly and loop geometries provides a remarkable five orders of magnitude improvement in fidelity. In the specific case shown, the linear geometry performs even better with infidelity reduced by seven orders of magnitude.

\section{Conclusion}
In summary, we have presented two-qubit dynamically decoupled gate pulse sequences in several different geometries of exchange-only qubit networks and most importantly, have shown how to make them resilient to quasistatic hyperfine nuclear spin noise. These sequences satisfy the physical constraints of the exchange-only system, representing dynamically corrected multiqubit gate operations applicable for exchange qubits. Our results, together with the corrected single-qubit gate pulse sequence in Ref.~\onlinecite{Hickman}, allow universal and robust multiqubit operations, which is an important step towards scalable fault-tolerant quantum computation on the exchange-only qubit platform in semiconductor quantum dots.

\section{Acknowledgments}
We thank Lev S. Bishop for stimulating discussions and Thaddeus D. Ladd for helpful correspondence. This work was supported by IARPA, LPS-CMTC, and JQI-NSF-PFC.

\appendix

\section{Comparison to single-pulse schemes for suppression of intrinsic leakage}\label{sec:doherty}

\begin{figure}[h!]
\centering
\includegraphics[ scale = 0.95] {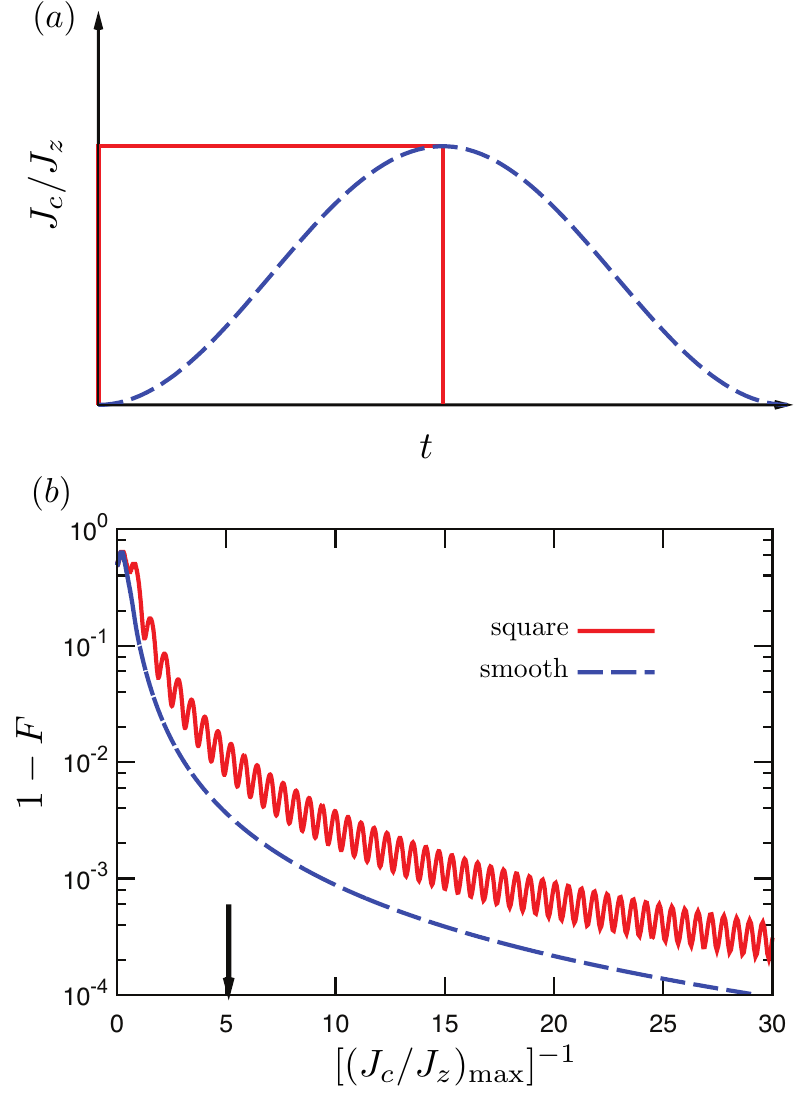}
\caption {(Color online) (a) Profile of the square pulse (red solid line) and the smooth pulse (blue dashed line) chosen to be proportional to $1-\cos2\pi t$. (b) Infidelity vs $[(J_c/J_z)_{\rm max}]^{-1}$. The $x$ axis is directly proportional to the gate duration for a square pulse according to Eq.~\eqref{suppleq:adiabaticpulse}, while the time duration for our smooth pulse is twice as long as that for the square pulse. The arrow marks the value of $J_c/J_z$ for the square pulse to have equal time duration of our pulse sequence.}\label{fig:adiabatic}
\end{figure}

In Ref.~\onlinecite{doherty}, it is shown that when intrinsic leakage is neglected, the two-qubit gate for a resonant-exchange qubit can be done in a relatively simple way, which, for the butterfly geometry, requires only one pulse. In order for the intrinsic leakage to be negligible, one must have the coupling exchange interaction $J_c$ very small compared to the typical exchange interaction $J_z$ for individual qubits $A$ (dots 1,2 and 3) and $B$ (dots 4,5 and 6), $J_c\ll J_z$, resulting in a slow gate. On the other hand, the sequences presented in our paper are completely immune to intrinsic leakage, but require a pulse sequence rather than a single pulse.  In this section, we discuss which method results in a faster gate for a given maximum tolerable infidelity.  We consider the butterfly geometry only since all other geometries require either local unitaries or echo sequences to compensate for the non-$\sigma_z\otimes\sigma_z$ rotations \cite{doherty}.

For the butterfly geometry considered in Ref.~\onlinecite{doherty}, the effective coupling Hamiltonian with all leakage states ignored reads
\begin{equation}
H_c=-\frac{J_z}{18}(\sigma_z\otimes I+I\otimes\sigma_z)+\frac{J_c}{9}\sigma_z\otimes\sigma_z.
\end{equation}
(Here it has been assumed that the intraqubit exchange interactions for both qubits $A$ and $B$ are equal, $J_{zA}=J_{zB}=J_z$.) We consider two types of pulses, a square pulse and a smooth pulse. The smooth pulse is chosen to be proportional to $1-\cos2\pi t$ and we enforce that its maximal $J_c$ value coincides with the one used in the square pulse. As a result, the gate corresponding to our smooth pulse is twice as long as the square pulse. See Fig.~\ref{fig:adiabatic}(a) for profiles of these pulses.

The time required to have a {\sc cphase} would be $9\pi/4 J_c$ for a square pulse and $2\times9\pi/4 J_c$ for our choice of smooth pulse. We show for both cases, the infidelity as a function of $[(J_c/J_z)_{\rm max}]^{-1}$ in Fig.~\ref{fig:adiabatic}(b). (See Appendix \ref{sec:fidel} for the formula used to calculate the fidelity.) The maximum ratio $(J_c/J_z)_{\text{max}}$ for any infidelity tolerance can be read off the figure and the minimum time for the gate is given by
\begin{equation}
T_{\text{min, 1 piece}}=\frac{9\pi}{4 J_{\text{max}} (J_c/J_z)_{\text{max}}},\label{suppleq:adiabaticpulse}
\end{equation}
where $J_{\text{max}}$ is the largest experimentally accessible value of the exchange coupling.

We now turn to the sequence in our work. Note that our pulse sequences are perfectly protected from intrinsic leakage, so the corresponding infidelity is always zero.  For the sequence presented in Fig.~\ref{fig:butterfly} appropriate for the butterfly geometry, one rotates by a total of $23\pi/2$.  The minimum time for our realization of the gate is then
\begin{equation}
T_{\text{min, sequence}}=\frac{23\pi}{2 J_{\text{max}}}.
\end{equation}
Thus, our pulse sequence is faster if $(J_c/J_z)_{\text{max}} < 9/46$ [marked as an arrow in Fig.~\ref{fig:adiabatic}(b)], which corresponds to a critical infidelity on the order of $10^{-3}$ for the square pulse and for the smooth pulse with twice the duration. Although it should be possible to further optimize the shape of the smooth pulse to reduce the infidelity, we note here that the pulse sequences discussed in this work go further than correcting intrinsic leakage: They also provide a systematic method to correct the hyperfine noise. It is unclear how one could achieve the same goal with a smooth pulse, and in cases where noise is significant one should use the corrected sequences we present despite their complexity compared to single-piece pulses such as discussed in this section.

\section{Remarks on the exhaustive search algorithm}\label{sec:algorithm}
In this section we make a few remarks on the optimization of the exhaustive search algorithm discussed in Sec.~\ref{sec:intrinsic}. To reduce the computational cost, we limit our search to time-order symmetric pulse sequences, in the sense that the same sequence is obtained when the order of the execution of the pulses is reversed. By restricting ourselves to this kind of sequences, we need to search only for the first half of the {\sc cnot} sequence as the second half can be constructed by simply reversing the order of the pulses in the first half. Since our goal is a {\sc cnot} gate up to local unitaries, we begin constructing the sequence by utilizing an inter-qubit pulse, for otherwise the first pulse could be absorbed into the local unitary gate. Subsequently, the construction of the sequence proceeds by restricting the search at each time step to  those pulses having the property that there exists at least one pulse in the previous time step that does not commute with it. This is to ensure that we do not end up with multiple sequences that are equivalent to each other by virtue of commuting pulses being able to be executed in any order.

Starting from the second time step, the sequence at each time step derived by using the above method is checked as to whether it is locally equivalent to a {\sc cnot} gate. The verification is done by first combining the sequence derived up to current time step with its reverse sequence i.e., the sequence obtained by reversing the execution order of the pulses leading to the current time step. Subsequently, the two invariants $G_{1,2}$ \cite{Makhlin} are calculated for this combined sequence. The sequence will be accepted if it has the same invariants as {\sc cnot}, otherwise it is rejected and another pulse combination satisfying the above constraints are tried and checked. This process continues until all possible pulse combinations from the first until the current time step, used to construct the first-half of the sequence, have been exhausted. If no solution is found, the construction of the sequence proceeds by adding another time step and checking the sequence derived from all possible pulse combinations. This process is carried on until a {\sc cnot} sequence is found.

\section{Implementation for linear, loop and other geometries}\label{sec:linear}

\begin{figure*}[h!]
\includegraphics[scale = 0.8] {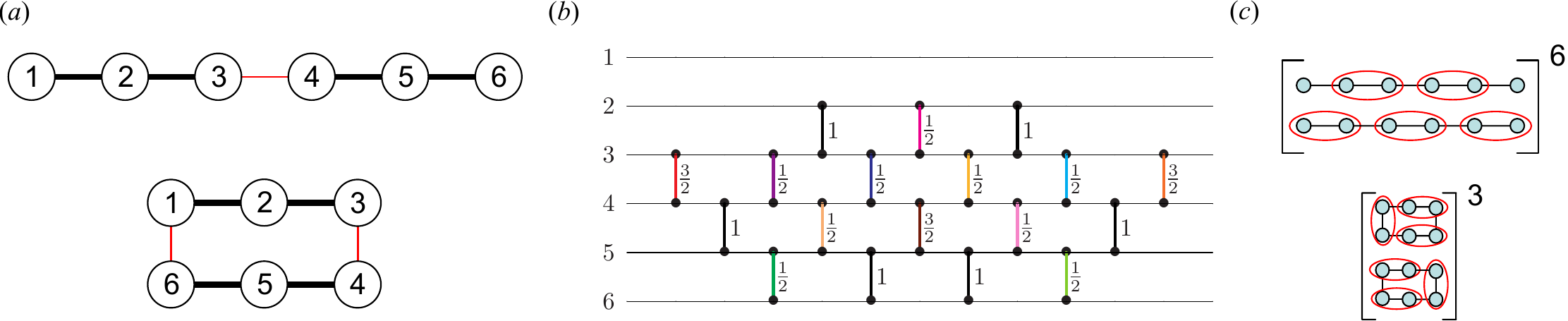}
\caption {(Color online) (a) Linear (top) and loop (bottom) geometries. (b) Schematic diagram of 18 exchange pulses in 11 time steps that realize a {\sc cnot} gate (up to local unitaries).\cite{Fong} The color coding used in this and subsequent diagrams should be interpreted in the same way as that in Sec.~\ref{sec:intrinsic}. (c) Sequences of complete cycles of permutations for linear (top) and loop (bottom) geometries.}\label{fig:linear}
\end{figure*}

\begin{figure*}[h!]
			\includegraphics[scale = 0.8]{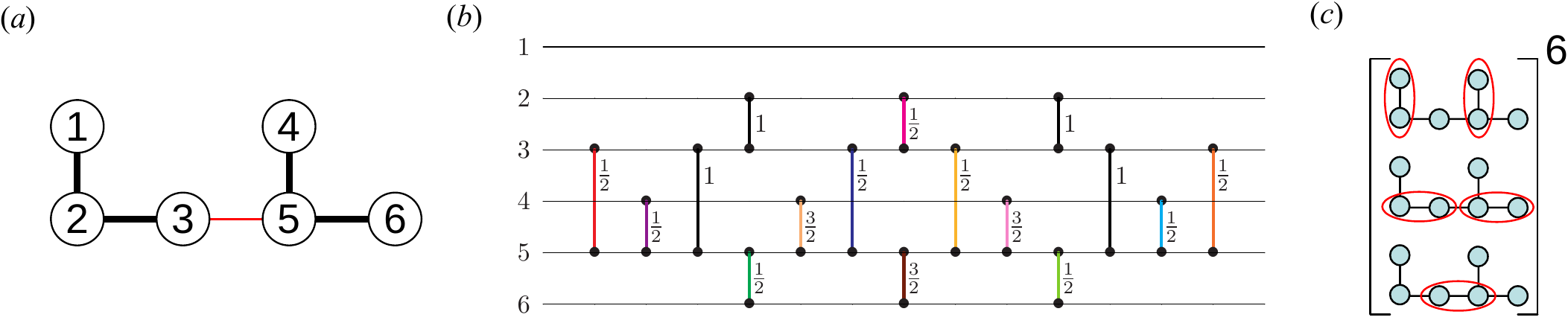}	
	\caption {(Color online) (a) Comb geometry. (b) Schematic diagram of 16 exchange pulses in 13 time steps
that realize a {\sc cnot} gate (up to local unitaries). (c) Sequence of complete cycle of permutations.}\label{fig:Lnetwork}
		\end{figure*}

\begin{figure*}[h!]
			\includegraphics[scale = 0.8]{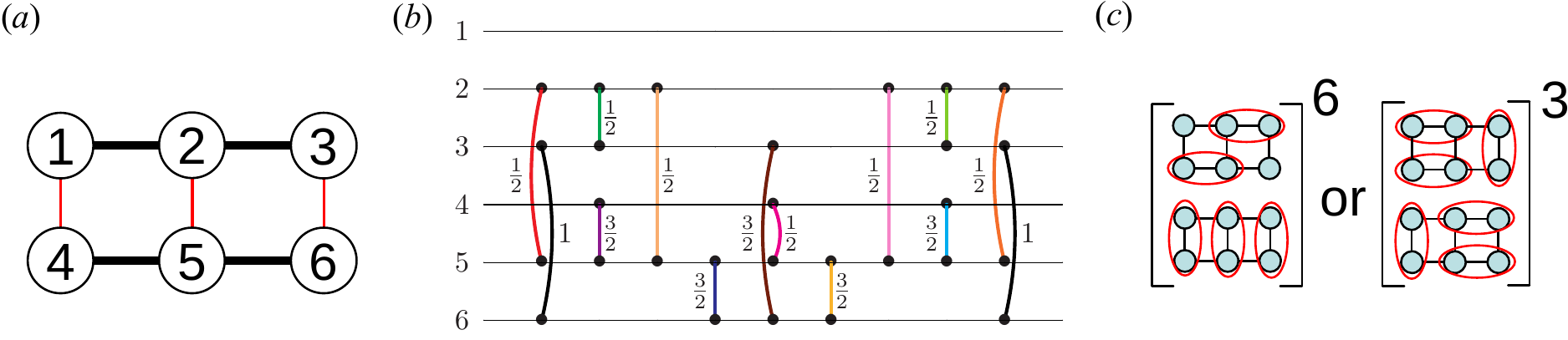}			
	\caption{(Color online) (a) Rectangular geometry in Ref.~\onlinecite{doherty}. (b) Schematic diagram of 14 exchange pulses in 9 time steps that realize a {\sc cnot} gate (up to local unitaries).  Note that at the first time step, rotations around axes 14 and 25 are to be carried out simultaneously. This bending of the links should be understood in the same way in the entire paper.  (c) Sequences of complete cycles of permutations.}\label{fig:ladder}
		\end{figure*}

\begin{figure*}[h!]
			\includegraphics[scale = 0.8]{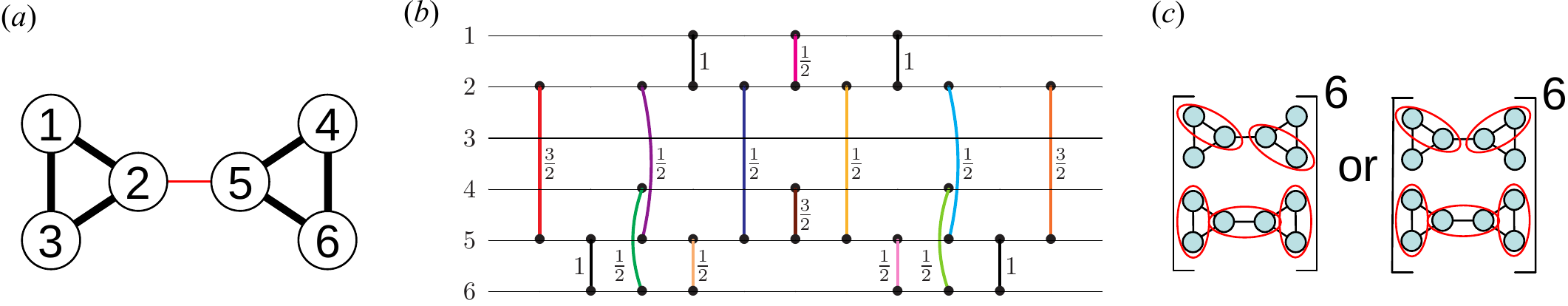}			
\caption{(Color online) (a) Bowtie geometry. (b) Schematic diagram of 16 exchange pulses in 11 time steps that realize a {\sc cnot} gate (up to local unitaries).(c) Sequences of complete cycles of permutations.}\label{fig:bowtie}
\end{figure*}

\begin{figure*}[h!]
			\includegraphics[scale = 0.8]{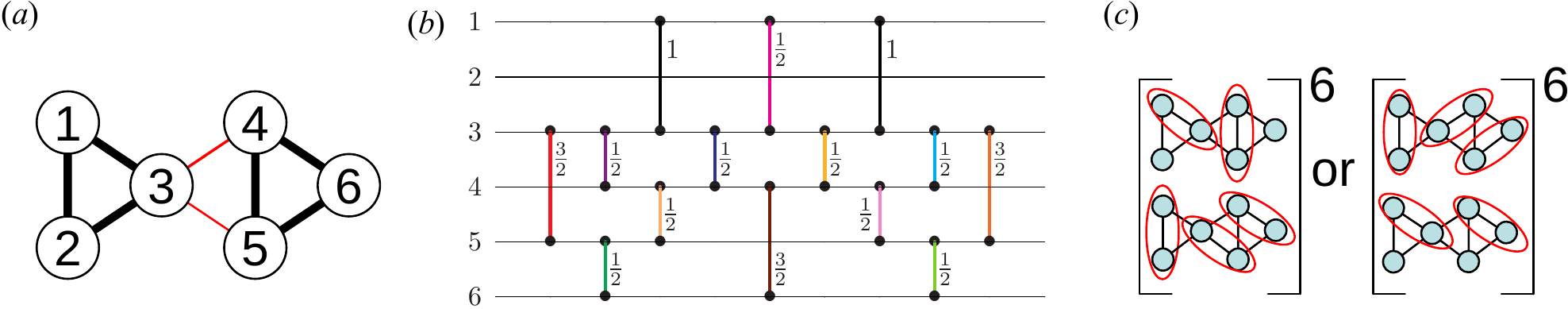}
\caption{(Color online) (a) One of the geometries considered in Ref.~\onlinecite{zhan}. (b) Schematic diagram of 14 exchange pulses in 9 time steps that realize a {\sc cnot} gate (up to local unitaries). (c) Sequences of complete cycles of permutations.}\label{fig:fish}
\end{figure*}

\begin{figure*}[h!]
			\includegraphics[scale = 0.8]{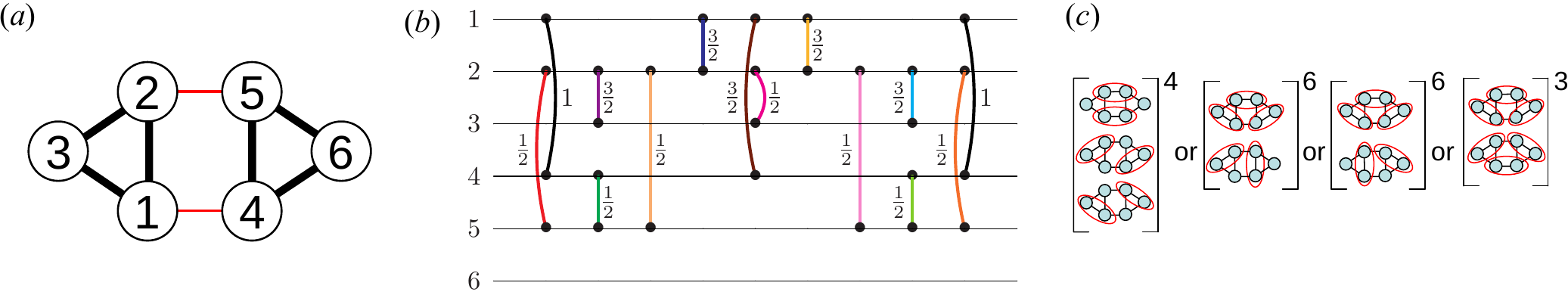}			
\caption{(Color online) (a) One of the geometries considered in Ref.~\onlinecite{zhan}. (b) Schematic diagram of 14 exchange pulses in 9 time steps that realize a {\sc cnot} gate (up to local unitaries). (c) Sequences of complete cycles of permutations.}\label{fig:HSBC}
	\end{figure*}

In this section we give {\sc cnot} sequences and permutation schemes that correct the leading hyperfine error term in different geometries (see Figs. \ref{fig:linear} - \ref{fig:HSBC}). In each of these figures, panel (a) shows the geometry with dots labeled, panel (b) shows the (locally equivalent) {\sc cnot} sequences, and panel (c) shows the cyclic permutation that one can implement by using $U_{ij}'(J,\phi)$ as the {\sc swap} operation to obtain a hyperfine-corrected sequence. For all the {\sc cnot} sequences given in panel (b), the exchange times are expressed in units such that for $t=1$, the rotation $\mathrm{exp}(-\imath t J {\bf{S}}_i\cdot{\bf{S}}_j)$ corresponds to a full {\sc swap} operation between spins $i$ and $j$.

Let us first consider the noise-correction sequence for linear and loop geometries. Understanding the permutation schemes for these two geometries aids in constructing the noise-correction procedure for the other geometries considered in this section. That is why it is worth going through the details of the schemes for these two geometries even though the general idea needed to construct the sequences has already been presented in Sec.~\ref{sec:hyperfine}.

The simplest and most popular geometry is a linear network of six quantum dots as shown in the top diagram of Fig.~\ref{fig:linear}(a). The sequence compensating intrinsic leakage was first presented in Ref.~\onlinecite{Divincenzo} and subsequently optimized by Fong and Wandzura \cite{Fong}, shown here in Fig.~\ref{fig:linear}(b).  In this geometry, the exchange coupling terms inside the set $E_A = \{E_{12},E_{34},E_{56}\}$ commute with each other and so do the coupling terms in $E_B = \{E_{23},E_{45}\}$ [see Fig.~\ref{fig:linear}(c)]. To swap the electrons in neighboring dots, we use the three-piece pulses $U'_{A}(J,\pi) \equiv U'_{12}(J,\pi)U'_{34}(J,\pi)U'_{56}(J,\pi)$ and $U'_{B}(J,\pi) \equiv U^{\mathrm{free}}_{16}(4\pi/J)U'_{23}(J,\pi)U'_{45}(J,\pi)$. Using these swapping pulses, we can construct a hyperfine-corrected identity by swapping the electrons in a way given in the top diagram of Fig.~\ref{fig:linear}(c), which gives
\begin{align}\label{eq:identity}
[U'_A(J,&\pi)U'_B(J,\pi)]^6\nonumber\\
&= \mathrm{exp}\left[-\imath\left(\frac{\pi}{2} + 8\pi \sum_{i=1}^6\frac{E^{\mathrm{Z}}_i}{J}\right)\right] I+ O\left((E^{\mathrm{Z}}_i/J)^2\right).
\end{align}

To implement a corrected simultaneous rotation about the three commuting axes (12, 34, and 56), we replace the last pulse $U'_A(J,\pi)$ of the corrected identity $[U'_A(J,\pi)U'_B(J,\pi)]^6$ by $U'_{12}(\pi+\phi_{12})U'_{34}(\pi+\phi_{34})U'_{56}(\pi+\phi_{56})$. The total sequence that realizes a corrected rotation simultaneously about axes 12, 34, and 56 can be written as
\begin{multline}
U'_{12}(\pi+\phi_{12})U'_{34}(\pi+\phi_{34})U'_{56}(\pi+\phi_{56})U_{B}'(J,\pi)\\
\times[U'_A(J,\pi)U'_B(J,\pi)]^5 \\
=\mathrm{exp}\left[-\imath\left(\frac{\pi}{2} + 8\pi \sum_{i=1}^6\frac{E^{\mathrm{Z}}_i}{J}\right)\right] R_{12}(\phi_{12})R_{34}(\phi_{34})R_{56}(\phi_{56})\\
+ \mathcal{O}[(E^{\mathrm{Z}}_i/J)^2].\label{eq:linear}
\end{multline}

This corrected sequence can be used for example to correct the hyperfine error in the third time step of the {\sc cnot} sequence shown in Fig. \ref{fig:linear}(b) with the angles in Eq. \eqref{eq:linear} being $\phi_{12} = 0$, $\phi_{34} = \pi/2$, and $\phi_{56} = \pi/2$.

Similarly, a corrected simultaneous rotation about axes 23 and 45 [e.g. the fourth time step in Fig.~\ref{fig:linear}(b)] can be obtained by replacing the last pulse $U'_B(J,\pi)$ of the corrected identity sequence $[U'_{B}(J,\pi)U'_{A}(J,\pi)]^6$ by $U^{\mathrm{free}}_{16}(4\pi/J)U'_{23}(J, \pi+\phi_{23})U'_{45}(J,\pi+\phi_{45})$. The complete sequence is given by

\begin{multline}
U^{\mathrm{free}}_{16}(4\pi/J)U'_{23}(J, \pi+\phi_{23})U'_{45}(J,\pi+\phi_{45}) U_{A}'(J,\pi)\\
\times [U'_B(J,\pi)U'_A(J,\pi)]^5\\
= \mathrm{exp}\left[-\imath\left(\frac{\pi}{2} + 8\pi \sum_{i=1}^6\frac{E^{\mathrm{Z}}_i}{J}\right)\right] R_{23}(\phi_{23})R_{45}(\phi_{45})\\
+ O((E^{\mathrm{Z}}_i/J)^2).
\end{multline}

A close variant of the linear geometry is a loop configuration as depicted in the bottom diagram of Fig.~\ref{fig:linear}(a). The only difference between these two geometries is the extra degree of freedom to exchange couple the electrons in dots 1 and 6 for the loop geometry. In fact, the {\sc cnot} sequence in Fig.~\ref{fig:linear}(b) developed for the linear geometry works seamlessly for the loop geometry. Due to the additional link between dots 1 and 6, the hyperfine-corrected sequence for the loop geometry has half the length of that for the linear geometry, as detailed below. 

To correct for the hyperfine noise in this geometry, we use the pulses $U'_{A}(J,\phi) \equiv U'_{12}(J,\phi)U'_{34}(J,\phi)U'_{56}(J,\phi)$ and $U'_B(J,\phi) \equiv U'_{16}(J,\phi)U'_{23}(J,\phi)U'_{45}(J,\phi)$. In this network, a complete cycle of permutations $[U'_A(J,\pi)U'_B(J,\pi)]^3$ gives $\exp\left[\imath\left(\pi/2-4\pi\sum_{i=1}^6{E^{\mathrm{Z}}_i}/{J}\right)\right]I+\mathcal{O}((E^{\mathrm{Z}}_i/J)^2)$. Similar to the case of the linear network, the corrected simultaneous rotation about axes 12, 34, and 56 is performed by replacing the last swapping pulse $U'_A(J,\pi)$ in the permutation sequence $[U'_A(J,\pi)U'_B(J,\pi)]^3$ by $U'_{12}(J,\pi+\phi_{12})U'_{34}(J,\pi+\phi_{34})U'_{56}(J,\pi+\phi_{56})$. On the other hand, for the corrected simultaneous rotation about axes 16, 23, and 45, the composite pulse $U'_{16}(J,\pi+\phi_{16})U'_{23}(J,\pi+\phi_{23})U'_{45}(J,\pi+\phi_{45})$ is used to replace the last swapping pulse in $[U'_B(J,\pi)U'_A(J,\pi)]^3$.

The permutation schemes derived for butterfly, loop, and linear geometries can also be utilized for other geometries. Consider the comb geometry shown in Fig.~\ref{fig:Lnetwork}. This geometry has the advantage, for example, to host either exchange-only qubits or the singlet-triplet qubit when the electrons inside dots 1 and 4 are depleted. The permutation scheme for this geometry [see Fig.~\ref{fig:Lnetwork}(c)] follows from that for butterfly geometry. On the other hand, the permutation schemes for the remaining geometries considered in this section (see Figs.~\ref{fig:ladder}-\ref{fig:HSBC}) follow the schemes for linear or loop geometries. For instance, the permutation sequence for the rectangular geometry shown in the left (right) part of Fig.~\ref{fig:ladder}(c) is the same as that for a linear (loop) geometry with the chain here being 1-4-5-2-3-6 (1-2-3-6-5-4-1). The results for the geometries considered in this section are presented in Figs.~\ref{fig:linear}-\ref{fig:HSBC}.

\section{Definition of fidelity}\label{sec:fidel}

The fidelity $F$ in Figs.~\ref{fig:infidelity} and \ref{fig:adiabatic} is computed using the formula~\cite{Kestner,Bowdrey02,Cabrera07}
\begin{align}\label{eq:Fidelity}
F  = & \frac{1}{16}\left[\frac{4}{5}\mathrm{Tr}[\sigma_0\otimes\sigma_0U_f\sigma_0\otimes\sigma_0U_f^\dagger]\right.  \nonumber\\
& \left. + \frac{1}{5}\sum_{\mu ,\nu = 0}^{3}\mathrm{Tr}[V\sigma_\mu\otimes\sigma_\nu V^\dagger U_f \sigma_\mu \otimes \sigma_\nu U_f^\dagger] \right],
\end{align}
where $\sigma_0$ is the 2 $\times$ 2 identity matrix and $\sigma_\mu \otimes \sigma_\nu$ are the Pauli matrices acting on the first and second qubits within the logical subspace (namely the upper left $4\times4$ blocks, with the remaining entries in the entire $9\times9$ matrix being zero), $U_f$ is the actual evolution of the composite pulse sequence, and $V$ is the desired operation with the identity in the leakage subspace.

\end{document}